\documentclass[aps,prl,twocolumn,superscriptaddress,groupedaddress]{revtex4}
\usepackage{subfig}
\usepackage{graphicx}  
\usepackage{dcolumn}   
\usepackage{bm}        
\usepackage{amssymb}   
\usepackage{slashed}
\usepackage{graphicx}				
\usepackage{amsmath}
\usepackage{mathtools}
\usepackage{tikz,pgf}
\usepackage{comment}
\usepackage{asymptote}
\usetikzlibrary{arrows,backgrounds}
\usetikzlibrary{fit,scopes,calc,matrix,positioning,decorations.pathmorphing}
\usepackage[all]{xy}
\usepackage{yfonts}
\newcommand{\bra}[1]{\ensuremath{\left\langle#1\right|}}
\newcommand{\ket}[1]{\ensuremath{\left|#1\right\rangle}}

\begin{document}
\title{Cosmological constant as quantum error correction from generalised gauge invariance in double field theory }
\author{Andrei T. Patrascu}
\address{ELI-NP, Horia Hulubei National Institute for R\&D in Physics and Nuclear Engineering, 30 Reactorului St, Bucharest-Magurele, 077125, Romania}
\begin{abstract}
The holographic principle and its realisation as the AdS/CFT correspondence leads to the existence of the so called precursor operators. These are boundary operators that carry non-local information regarding events occurring deep inside the bulk and which cannot be causally connected to the boundary. Such non-local operators can distinguish non-vacuum-like excitations within the bulk that cannot be observed by any local gauge invariant operators in the boundary. The boundary precursors are expected to become increasingly non-local the further the bulk process is from the boundary. Such phenomena are expected to be related to the extended nature of the strings. Standard gauge invariance in the boundary theory equates to quantum error correction which furthermore establishes localisation of bulk information. I show that when double field theory quantum error correction prescriptions are considered in the bulk, gauge invariance in the boundary manifests residual effects associated to stringy winding modes. Also, an effect of double field theory quantum error correction is the appearance of positive cosmological constant. The emergence of spacetime from the entanglement structure of a dual quantum field theory appears in this context to generalise for de-Sitter spacetimes as well.  
\end{abstract}
\maketitle
\section{Introduction}
The AdS/CFT correspondence and its underlying holographic principle lead to a new way of regarding quantum field theories and their observables. Moreover, spacetime together with its geometry is seen as an emergent property of the entanglement structure of the boundary conformal field theory. It has been noticed that a local bulk observable is not in a one to one relation to boundary operators. One bulk observable can be holographically represented in many different forms by operators of the boundary conformal field theory. If one considers any point $x$ in the bulk space and one takes a point $Y$ on the boundary, then holography, as implemented by the AdS/CFT correspondence can map an associated bulk local field $\phi(x)$ into many different possible CFT operators. One can consider $\mathcal{O}[\phi(x)]$ which has no support in an open set containing $Y$ and holographically map the local bulk field to it. As a consequence, any local field of the CFT supported near $Y$ will commute with it. As we considered $Y$ to be an arbitrary point, if the CFT operator corresponding to it were to be unique we would arrive at the conclusion that $\mathcal{O}$ commutes with all local fields in the CFT and hence must be a multiple of the identity as the local field algebra is irreducible. However, one does not expect that any local bulk field to be mapped into the identity operator on the boundary. Indeed, one can avoid this conclusion if one realises that the bulk-boundary correspondence is not one-to-one or unique. Considering $Y$ and $Z$ two distinct boundary points, then the bulk operator $\phi(x)$ may be mapped either to the operator $\mathcal{O}$ on the boundary which commutes with the CFT local fields supported near $Y$, or to $\mathcal{O}'$ which commutes with the CFT local fields supported near $Z$. $\mathcal{O}$ and $\mathcal{O}'$ are inequivalent operators defined within the boundary CFT although they can be used interchangeably for the description of bulk physics. If we consider a fixed time subregion $A$ of the the CFT we can define a subregion in the bulk $\mathcal{C}[A]$ such that for any point $x\in \mathcal{C}[A]$, the bulk quantum field theory ensures that any bulk local operator $\phi(x)$ can be represented in the CFT as some non-local operator on $A$. While locality is not present in the bulk, one may imagine that it emerges by some mechanism. Bulk non-locality may be associated to a string theoretical effect and hence, to better understand its emergence one has to consider an intermediate situation between bulk non-locality and bulk locality. It is argued here that such an intermediate situation may be described by means of double field theory in the bulk. This theory would be an effective field theory that retains certain aspects of string theory related especially to their extended nature. Together with momentum modes, double field theory also incorporates winding modes which rely on the extended nature of strings and which may have effects that vanish when only supergravity type effective field theories are being considered. How to properly express double field theory in the context of the AdS/CFT correspondence however is highly non-trivial. If one considers a given bulk point $x$, one notices that it can lie within distinct causal wedges corresponding to different boundary regions, and hence the bulk operator $\phi(x)$ can have different representations in the boundary CFT with different spatial support. Reference [7] associates the non-uniqueness of the CFT operators corresponding to the operator $\phi(x)$ in the bulk with the possibility that the bulk field represents a form of logical operator which preserves a code subspace of the CFT Hilbert space. With such an interpretation it results that bulk field protects the CFT code subspace against erasures of parts of the boundary. Let the boundary operator corresponding to $\phi(x)$ act on a subsystem of the CFT which is protected against erasure of the boundary region denoted $A^{c}$. Such an operator can be represented in the boundary CFT as having a support on $A$, which is the complement of the erased region. The AdS-Rindler reconstruction of $\phi(x)$ on the boundary region $A$ can then be considered to act as a correction for the erasure of $A^{c}$. For various choices of the portion we wish to erase, we obtain different reconstructions of the bulk field $\phi(x)$. The deeper inside the bulk the operators are, the better protected against erasures they become, meaning that a larger region needs to be erased to prevent their reconstruction while operators near the boundary can be erased more easily by removing a smaller part of the boundary. As showed in [17] the code subspace can be seen as the low energy sector of the conformal field theory which corresponds to a smooth dual classical geometry. All the boundary CFT operators are physical and also have a bulk interpretation. The logical operators are special ones which map low energy states to other low energy states. Identical logical actions can be realised by distinct CFT operators, because those operators act on the high energy CFT states from outside the code subspace differently, although they act on low energy states in the same way [17]. The AdS/CFT holographic duality is known to map a D+1 dimensional conformal field theory on a flat spacetime into a D+2 dimensional quantum gravity theory on an AdS space background. It is well known that the large-N limit of the conformal field theory corresponds to the classical limit of the corresponding gravity theory. This correspondence is related to the existence of an SO(D,2) conformal symmetry group associated to the quantum field theory on the boundary which is identical to the isometry group of the AdS space. Different, more general quantum field theories, without conformal symmetry may be dual to bulk gravity theories with different spacetime manifolds. It is interesting to observe that the bulk theory becomes in the high energy limit a theory of quantum gravity. String theory, as such a theory, should therefore play an essential role in the holographic interpretation of various processes in the bulk. However, while it is sufficiently clear that string theory is holographical [18] there is no clear representation of the holographic duality in terms of string theoretical effects within the bulk. The recent understanding of the AdS/CFT correspondence in terms of error correction codes led to speculations on how various error correction properties may be understood in terms of the quantum field theory on the boundary. One idea was that any gauge invariant state already possesses some form of non-local entanglement originating in the initial requirement that it satisfies certain gauge constraints [1]. The connection between gauge invariance in the boundary theory and the emergence of the bulk spacetime appears in the natural error correction code that imposes localisation of the bulk information in different regions. It appears that such a connection is more general than the standard one based on error correction schemes. However, this connection has not been expressed in its most general form, as it did not take into account properties that may appear due to winding modes in the bulk. Such modes are not included in standard interpretations, yet they alter the gauge invariance transformations in a significant way. If we could re-interpret those generalised gauge transformations in terms of certain quantum information properties, our understanding both of quantum gravity and of quantum computing would significantly increase. If gauge invariance implies a form of entanglement for the boundary quantum field theory, how can that be generalised following the extension of the standard gauge invariance transformations to those associated to double field theory? 
\section{Double Field Theory}
The introduction of gauge symmetries in quantum information theory has been recently explored in [1] and [2]. Quantum error correction codes were required because the classical error correction based on analysing copies of the same information for discrepancies cannot be applied in quantum computing due to the no-cloning theorem. A way of providing additional robustness and the ability of correcting potential errors in quantum codes was brought by the fact that entanglement encodes the required information globally. A similar situation occurred when analysing precursor operators on the boundary of a holographic theory. Events occurring deep within the bulk could not be causally related to the boundary, yet, due to holography, they were encoded within the boundary by means of non-local operators. These non-local operators are what we call precursors and they exploit the fact that local information within the bulk can be represented non-locally on the boundary. Moreover, it has been showed that error correction in the boundary is linked to the gauge invariance of the boundary theory. Another situation in which non-local phenomena manifest themselves is double field theory. As this theory incorporates the string theoretical T-duality [3] which connects different length scales and even different topologies, understanding quantum error correction in the context of double field theory in the bulk will be of particular relevance, increasing the theoretical resilience of quantum codes on the boundary even further. As non-trivial topology has been shown to be equivalent to entanglement at least in the bipartite case [4], [5], understanding the effect of T-duality on quantum error corrections could provide new insights on the sets of observables that can come together and provide information about the boundary (when local) and about the causally disconnected events in the bulk (when non-local).
However, in double field theory, gauge invariance is extended and the generalised metric is used. The theory we try to write in a T-duality invariant fashion is the NS-NS sector of supergravity. The degrees of freedom of this theory are contained in the $D$-dimensional metric tensor $g_{ij}$, $i,j=1,...,D$, the $D$-dimensional $2$-form $b_{ij}$ (Kalb-Ramond field) and the dilaton $\phi$, all depending on the spacetime coordinates $x^{i}$. Physical phenomena will not change under a pair of local gauge transformations. The first such symmetry is the diffeomorphism parametrised by the infinitesimal vectors $\lambda$ and encoded by the Lie derivative acting on arbitrary vectors $V^{i}$ like
\begin{equation}
L_{\lambda}V^{i}=\lambda^{j}\partial_{j}V^{i}-V^{j}\partial_{j}\lambda^{i}=[\lambda,V]^{i}
\end{equation}
where the last term is a Lie bracket which is antisymmetric and satisfies the Jacobi identity. The second gauge symmetry transformation of the $2$-form parametrised by the infinitesimal $1$-form parameter $\tilde{\lambda}_{i}$ is
\begin{equation}
b_{ij}\rightarrow b_{ij}+\partial_{i}\tilde{\lambda}_{j}-\partial_{j}\tilde{\lambda}_{i}
\end{equation}
The supergravity action takes the well known form 
\begin{equation}
S=\int d^{D}x\sqrt{g}e^{-2\phi}[R+4(\partial \phi)^{2}-\frac{1}{12}H^{ijk}H_{ijk}]
\end{equation}
where the three-form $H_{ijk}=3\partial_{[i}b_{jk]}$ satisfies the Bianchi identity $\partial_{[i}H_{jkl]}=0$ and $R$ is the Ricci scalar constructed from $g_{ij}$. 
In double field theory, these gauge symmetries are extended with the explicit addition of the T-duality symmetry which relates the fields $g_{ij}$, $b_{ij}$, and $\phi$ mentioned above. To see how supergravity degrees of freedom can be put in a T-duality invariant formulation let us arrange all the objects in T-duality representations, hence having well defined transformation properties with respect to T-duality [6]. The metric and the two-form field can be combined into the symmetric generalised metric 
\begin{equation}
\mathcal{H}=
  \left({\begin{array}{cc}
   g^{ij} & -g^{ik}b_{kj} \\
   b_{ik}g^{kj} & g_{ij}-b_{ik}g^{kl}b_{lj} \\
  \end{array} } \right)
\end{equation}
This metric is an $O(D,D)$ group element satisfying the property that its inverse is obtained by acting with the Minkowski metric on it
\begin{equation}
\eta_{MN}=
  \left({\begin{array}{cc}
   0 & \delta^{i}_{\;\; j} \\
   \delta_{i}^{\;\;j} & 0 \\
  \end{array} } \right)
\end{equation}
\begin{equation}
\mathcal{H}^{MN}=\eta^{MP}\mathcal{H}_{PQ}\eta^{QP}
\end{equation}
where the uppercase indexes go from $1$ to $2D$ and refer to the doubled space. One may employ the notation $\sqrt{g}e^{-2\phi}\rightarrow e^{-2\phi}$ transforming the dilaton term which now becomes an $O(D,D)$ scalar. Using the standard notation for doubled coordinates $X^{M}=(\tilde{x}_{i},x^{i})$ we have the new coordinates $\tilde{x}_{i}$ representing the coordinates associated to the winding modes of the strings. In the context of supergravity these coordinates have no meaning and hence we need some form of constraint that while being T-duality invariant must restrict the dependence on such coordinates. There are several such constraints, one of the most common being the so called section (or strong) constraint $\eta^{MN}\partial_{M}\partial_{N}(...)=0$. This will be assumed in what follows. Following the notation of [6] this will be rewritten as $Y^{M\;\;N}_{\;\;\;\; P\;\;\; Q}\partial_{M}\partial_{N}(...)=0$ where $Y^{M\;\;N}_{\;\;\;\; P\;\;\; Q}=\eta^{MN}\eta_{PQ}$. The metric $g_{ij}$ and the Kalb-Ramond field transform under diffeomorphisms and the Kalb-Ramond field also transforms under gauge symmetry. In double field theory the standard diffeomorphisms can be unified with the gauge transformation leading to a generalised diffeomorphism implemented by a generalised Lie derivative and encoding the generalised gauge transformations of the two entities. The gauge parameter can be written in the double field theory as
\begin{equation}
\xi^{M}=(\tilde{\lambda}_{i},\lambda^{i})
\end{equation}
and the resulting generalised Lie derivative acting on a tensorial density $V^{M}$ with weight $\omega(V)$ will have the form 
\begin{equation}
\mathcal{L}_{\xi}V^{M}=\xi^{P}\partial_{P}V^{M}+(\partial^{M}\xi_{P}-\partial_{P}\xi^{M})V^{P}+\omega(V)\partial_{P}\xi^{P}V^{M}
\end{equation}
Note that in this case $\omega(e^{-2\phi})=1$ and $\omega(\mathcal{H})=0$. The closure of these generalised diffeomorphisms imposes certain differential constraints on the theory. The closure of the group law implies that two successive gauge transformations parametrised by $\xi_{1}$ and $\xi_{2}$ acting on a given field $\xi_{3}$ generate a new transformation of the same group parametrised by $\xi_{12}(\xi_{1},\xi_{2})$ 

\begin{equation}
([\mathcal{L}_{\xi_{1}},\mathcal{L}_{\xi_{2}}]-\mathcal{L}_{\xi_{12}})\xi_{3}^{M}=0
\end{equation}
i.e. the generalised Lie derivative must send tensors to tensors. The parameter resulting from this is $\xi_{12}=\mathcal{L}_{\xi_{1}}\xi_{2}$ with the constraint 
\begin{equation}
Y^{M\;\;N}_{\;\;\;\; P\;\;\; Q}(2\partial_{P}\xi^{R}_{[1}\partial_{Q}\xi_{2]}^{M}\xi_{3}^{S}-\partial_{P}\xi_{1}^{R}\xi_{2}^{S}\partial_{Q}\xi_{3}^{M})=0
\end{equation}
The parameter $\xi_{12}$ is known as the D-bracket and its antisymmetric part (the C-bracket) is 

\begin{equation}
\xi_{[12]}^{M}=\frac{1}{2}(\mathcal{L}_{\xi_{1}}\xi_{2}^{M}-\mathcal{L}_{\xi_{2}}\xi_{1}^{M})=[\xi_{1},\xi_{2}]^{M}+Y^{M\;\;N}_{\;\;\;\; P\;\;\; Q}\cdot\xi_{[1}^{Q}\partial_{P}\xi_{2]}^{N}
\end{equation}

This corresponds to the extension of the usual Lie bracket, due to the correction depending on $Y$. This factor measures the departure from the conventional Riemannian geometry. 

 As can be seen, the explicit inclusion of T-duality gives rise to an extended form of gauge transformation which leads not only to a generalised Lie derivative and bracket but also to a new way of interpreting the quantum error correction and quantum secret sharing procedures in quantum gravitational contexts. 
Indeed, up to now quantum error correction was dominated by the fact that entanglement on the boundary encoded the bulk information non-locally with respect to boundary coordinates. Here, due to the introduction of coordinates related to the winding modes of the string, quantum error correction can make use of the additional $\tilde{x}$ coordinates as well. They are a stringy feature that allows us to use analogues of left and right rotation projectors in the boundary theory. Moreover, the patching of the boundary space will be somewhat unusual, as the patching function will now be related to the symmetry transformation defined by T-duality. This will result in a non-geometric structure on the boundary.
 It has been noted in [1] that while a local bulk operator is dual to several boundary precursor operators, those operators are all equivalent when acting on gauge invariant states. Given a bulk operator, the demand for bulk locality implies that the boundary precursor commutes with all spacelike separated boundary operators. Moreover, a bulk operator can be represented either as boundary precursors written in the form of bilocal operators distributed over the whole boundary or by representing bulk operators in the right bulk Rindler wedge as precursors smeared over the entire right half of the boundary [1]. This last representation allows one to eliminate bilocals connecting the two halves of the boundary or those who stretch only inside the left half of the boundary without any physical consequences. Such freedom arises because precursors may only act on gauge invariant states. The non-local nature of precursors relates to the idea of entanglement by the observation that given three patches $A$, $B$, and $C$ which cover the boundary conformal field theory, the precursors can only be reconstructed by combining at least two of these patches, i.e. $AB$, $BC$, or $CD$ but not from each $A$, $B$, or $C$ alone. This non-local storage of bulk information reminds us of entanglement and quantum error correction codes [7-9]. When double field theory is considered within the bulk we must remember that the symmetry made manifest by it is T-duality, which is specific to string theory. In the bulk this means we have to consider specific projector operators which single out left and right rotational states, a feature resulting from the closed string origins of this analysis. In the boundary limit this extends the way in which patches can be combined to provide useful information about the bulk states leading to information about the bulk state being encoded in combination of patches related via T-duality symmetry and resulting into non-geometric structures. Indeed a string can wrap around non-trivial cycles of the background leading to so-called winding states. Such states are created by vertex operators which depend on both coordinates associated with momentum excitations and T-dual coordinates associated with the winding excitations. Excitations of the vacuum by these operators may lead to non-geometric backgrounds. These backgrounds correspond to field theories with interactions depending on both types of coordinates. Given closed string theory in D-dimensional space with $d$ compactified directions, $\mathbb{R}^{n-1,1}\times T^{d}$ where $n+d=D$ and the coordinates $x^{i}=(x^{\mu},x^{a})$, $i=0,...,D-1$ where $a$ refers to the $d$-torus, the states are labelled by the momentum $p_{i}=(k_{\mu},p_{a})$ and the string windings $w^{a}$. States within the bulk will then be written as  
\begin{equation}
\ket{\Phi}=\sum_{I}\int dk\sum_{p_{a},w^{a}}\phi_{I}(k_{\mu},p_{a},w^{a})\mathcal{O}^{I}\ket{k_{\mu},p_{a},w^{a}}
\end{equation}
By means of a Fourier transform the dependence on the momenta is transformed into the spacetime dependence $x^{\mu}$ and $x^{a}$ while the winding mode $w^{a}$ becomes a new periodic coordinate which we called $\tilde{x}_{a}$. Physical strings will be annihilated by 
\begin{equation}
L_{0}-\bar{L}=N-\bar{N}-p_{a}w^{a}=0
\end{equation}
Fields in the double field theory can be extended to the boundary and in this limit we can express them in terms of boundary operators depending on both normal and winding coordinates. Keeping the generalised gauge invariance and writing the states in the bulk within the context of double field theory we may connect bulk operators to boundary precursors by first using the bulk equations of motion for $\ket{\Phi}$, express it in terms of the boundary fields, and then use the boundary equations of motion to evolve this to a single time operator [1]. Here, the gauge invariance of the bulk-space modes is larger and involves both double coordinates. Demanding closure of the group law does not necessarily restrict us to the normal supergravity bulk. Instead we may obtain non-geometric structures unreachable from supergravity alone. Indeed, gauge invariance constraints on the boundary are translated into quantum error correction prescriptions as predicted by [1] but the gauge invariance here incorporates naturally diffeomorphisms and gauge transformations of the generalised metric and 2-form field
\begin{widetext}
\begin{equation}
\begin{array}{c}
\mathcal{L}_{\xi}e^{-2\phi}=\partial_{M}(\xi^{M}e^{-2\phi})\\
\\
\mathcal{L}_{\xi}\mathcal{H}_{MN}=L_{\xi}\mathcal{H}+Y^{R\;\;M}_{\;\;\;\; P\;\;\; Q}\partial^{Q}\xi_{P}\mathcal{H}_{RN}+Y^{R\;\;N}_{\;\;\; P\;\;\; Q}\partial^{Q}\xi_{P}\mathcal{H}_{MR}
\end{array}
\end{equation}
\end{widetext}
where $L_{\xi}$ is the usual Lie derivative in $2D$ dimensions. The term $Y$ incorporates non-Riemannian effects. The requirement of gauge invariance to this type of transformations can be interpreted as a quantum error correction code built naturally into the precursor operators only that now the extension towards the double field theory bulk space gauge transformations not only encode that information is non-locally spread over the boundary but also that they have a non-geometric component. The closure requirement for such transformations when analysed in a doubled bulk spacetime allows non-geometric effects to participate to the error correction mechanism. Non-geometric effects represent a departure from strict Riemannian geometry and are fundamentally invisible from the perspective of local standard quantum field theory not involving winding coordinates. It has been shown in [10] and [11] that in double field theory the Riemannian tensor is not fully determined in terms of physical fields. Moreover, the components of the Riemannian tensor that do not contain undetermined connections are zero. We can however define a set of projectors 
\begin{equation}
\begin{array}{cc}
P_{M}^{\;\;\;N}=\frac{1}{2}(\delta_{M}^{\;\;\;N}-\mathcal{H}_{M}^{\;\;\;N}), & \bar{P}_{M}^{\;\;\;N}=\frac{1}{2}(\delta_{M}^{\;\;\;N}+\mathcal{H}_{M}^{\;\;\;N})
\end{array}
\end{equation}
which allow us to project onto the left-handed and right-handed subspaces. 
We use the simplifying notation for their action on indices as in [10]
\begin{equation}
\begin{array}{cc}
W_{\underline{M}}=P_{M}^{\;\;\;N}W_{N}, & W_{\bar{M}}=\bar{P}_{M}^{\;\;\;N}W_{N}
\end{array}
\end{equation}
The indeterminacy of the Riemann tensor can be traced back to the connection which can be decomposed into a determined and an undetermined part 
\begin{equation}
\Gamma_{MNK}=\hat{\Gamma}_{MNK}+\Sigma_{MNK}
\end{equation}
where the hat denotes the part of the connection determined by the physical fields while the undetermined part can again be decomposed as 
\begin{equation}
\Sigma_{MNK}=\tilde{\Gamma}_{\underline{M}\;\underline{N}\;\underline{K}}+\tilde{\Gamma}_{\bar{M}\bar{N}\bar{K}}
\end{equation}
To be sure that we work with a meaningful Riemannian curvature we have to define it in terms of projected indexes when working in double field theory, namely $R_{\underline{M}\;\underline{N}\;\underline{P}\;\underline{K}}$ and the scalar curvature then becomes 
\begin{equation}
R=R^{\underline{M}\;\underline{N}}_{\;\;\;\;\;\;\;\;\underline{M}\;\underline{N}}
\end{equation}
When this method is used the undetermined connections drop out of $R$ [10]. However, there always exists the additional freedom given by the indeterminacy of the connection. In terms of quantum error correction codes, this allows a spread of quantum information on non-geometric structures that in the boundary limit becomes not only inaccessible to any local observables but also inaccessible to observers who ignore the stringy structure given by the winding coordinates in double field theory. The cosmological constant term in the doubled approach is 
\begin{equation}
\int dy^{2D}e^{-2\phi}\Lambda
\end{equation}
with $\Lambda=\frac{4}{\alpha}$ as computed in [12] is required to match the DFT action with the effective action from string theory. This fact is particularly representative as normal effective actions do not encode non-geometric backgrounds. DFT is the first theory capable of detecting non-geometry. As has been shown before, extending gauge invariance to the generalised gauge invariance induced by double field theory we obtained additional tools for implementing quantum error correction, involving not only non-locality on Riemannian geometry but also spreading of the precursor operators in the boundary on manifolds defined by patching functions involving T-duality symmetry. This leads to non-Riemannian effects and to non-geometry. Non-localisation in non-geometry must take into account the way in which patches of the boundary manifold connect and T-duality plays a major role in this. Entanglement entropy has been associated with the cosmological constant in [13] yet bringing together the role of entanglement in quantum error correction codes has only recently been done in [14]. Connecting quantum error correction to gauge invariance has been done in [1] while finally in this work I relate the generalised gauge invariance of double field theory with an extended way of interpreting natural quantum error correction involving string-geometry phenomena. This generalised form of entanglement may be the source of the entanglement entropy part which plays a role in the cosmological constant. 

\section{Quantum error correction and stabiliser codes}
Recently, holography and its most direct manifestation, the AdS/CFT duality, have been interpreted in terms of quantum error correction protocols. In this sense, the quantum field theoretical description on the boundary, with its far larger number of degrees of freedom is regarded as the physical encoding of a logical quantum state manifested within the geometry of the bulk. To proper understand this interpretation and to relate T-duality and double field theory in the bulk with new forms of quantum error correction codes it is essential to understand the quantum error correction prescriptions in more detail. In general an arbitrary state of an individual qubit can be expressed as 
\begin{equation}
\ket{\phi}=\alpha\ket{0}+\beta\ket{1}
\end{equation}
with the two orthonormal basis states $\ket{0}$ and $\ket{1}$ and the coefficients satisfying $|\alpha |^{2}+|\beta |^{2}=1$. In quantum computation, the gate operations are represented by unitary operators acting on the Hilbert space of a collection of qubits. All operations must be reversible and hence unitary. The dynamical operation of a gate on a qubit is a member of the unitary group $U(2)$, $G$ which is a unitary matrix of dimension two such that $G^{\dagger}=G^{-1}$. Ignoring a global and unphysical phase factor, any gate operation on a qubit may be expressed as a linear combination of generators of the group $SU(2)$ in the form 
\begin{equation}
G=c_{I}\sigma_{I}+c_{x}\sigma_{x}+c_{y}\sigma_{y}+c_{z}\sigma_{z}
\end{equation}
where $\sigma_{i}$ are the Pauli matrices including the identity. The main difference between classical and quantum error correction lies in the fact that we cannot duplicate quantum states (no-cloning theorem) and we cannot directly measure a single quantum state without destroying its quantum nature. Therefore, error correction protocols must be adapted in order to detect and correct errors without being forced to acquire any information about the state itself. Qubits employed in quantum information are susceptible to the classical bit errors like bit switching, but also to phase errors. Hence quantum error correction must take into account both. Errors in quantum mechanics are inherently continuous, as qubits experience angular shifts of the qubit state by any possible angle. What quantum error correction has in common with classical error correction is its reliance on redundancy in the encoding prescription. Such redundancy implies that a single quantum state is encoded over a larger Hilbert space, extending the domain of representation of, say, a qubit, beyond what would be required for a single qubit. Extending the space of states in order to obtain auxiliary symmetries that could simplify certain computations has been used in ref. [2] and [19]. There, the extension was based on the Batalin-Vilkovisky quantisation of gauge theories with non-closing gauge algebras and the extensions in the form of field-anti-field formalisms [20], [21]. Here, the extension will at first play a different role, as the focus will be on quantum error corrections. However, it has been noted in [1] that gauge invariance in the boundary field theory may be related to quantum error correction. As there are various ways to implement gauge invariance and to construct meaningful quantum gauge theories, even in the case of non-closing algebras, it is interesting to see how this may relate to the construction of more efficient quantum error correction codes. Ultimately, the gauge invariance of double field theory, with its manifest T-duality symmetry imposes a set of relations valid for all scales and fundamentally non-local. It will be seen in this article that such non-local relations, connecting even distinct topologies, may be obtained from holographic quantum codes by allowing certain extensions with respect to the requirements of [17]. Before we discuss those connections and start constructing more advanced holographic quantum error correction codes, let me first describe what types of quantum errors are to be expected in any general quantum code. Surely, errors existing in any quantum system depend on the specific physical mechanisms controlling the system. In general however we can identify three types of errors: coherent quantum errors, due to incorrect application of quantum gates, environmental decoherence errors due to the interaction of the quantum system with the environment, and loss, or quantum leakage. In our situation the focus will be on coherent quantum errors and errors due to qubit erasure (or loss). Their correction relies on multiple qubit encoding of a single qubit quantum information and on correction of individual errors. A first simple example is the so called 3-qubit code, which, while not capable of simultaneously correcting both bit and phase flips, is one of the first repetition codes used finally by Shor [8] to construct the 9-qubit code capable of simultaneous bit and phase flip error corrections. The main idea of the 3-qubit code is to encode a single logical qubit into three physical qubits such that any single $\sigma_{x}$ bit flip error will be corrected. There will be two logical basis states defined in terms of three physical qubits: $\ket{0}_{L}=\ket{000}$ and $\ket{1}_{L}=\ket{111}$. In general an arbitrary qubit state can be reformulated as 
\begin{equation}
\begin{array}{c}
\alpha\ket{0}+\beta\ket{1}\rightarrow \alpha\ket{0}_{L}+\beta\ket{1}_{L}=\\
=\alpha\ket{000}+\beta\ket{111}=\ket{\psi}_{L}\\
\end{array}
\end{equation}
A quantum circuit that would encode such a state with three qubits will start with three quantum states, the first encoding the original qubit state, and another two ancilla qubits initialised to $\ket{0}$. Two CNOT gates will couple the first qubit state to the second $\ket{0}$ state and the second $\ket{0}$ state to the third such that, in the end, the logical qubit will be encoded on three qubits. This code features a binary distance between the two codeword states and hence is capable of correcting for a single bit flip error. It is necessary to have three physical bit flips in order to transform the logical state from $\ket{0}_{L}$ to $\ket{1}_{L}$. Therefore if we assume $\ket{\psi}=\ket{0}_{L}$, then with one single bit flip we will obtain a final state that still remains closer to $\ket{0}_{L}$. The distance between two codeword states, $d$, is related to the number of errors that can be corrected, $t$, by means of the relation 
\begin{equation}
t=[\frac{d-1}{2}]
\end{equation}
The error correction prescription on the other side will need some additional ancilla qubits, because we cannot directly measure the logical state without destroying it. Those ancilla qubits are used to extract the syndrome information related to possible errors without discriminating the state of any qubit. The error correction connects the physical qubits to the new ancilla qubits by means of CNOT gates which check the parity of the three-qubit data block. In any case, there is either no error, or a single bit-flip error and in both cases the ancilla qubits are flipped to one unique state based on the parity of the data block. These qubits are then measured and provide the syndrome of the error. This will then allow us to apply the correction gate in a meaningful way. In order to correct for both bit and phase flip, the nine-qubits code may be employed. Other generalisations are possible but the simple discussion up to this point suffices for the matter at hand. Describing error correction codes from the perspective of the quantum state is often cumbersome and inefficient as the state representations and the circuits themselves will differ from code to code. The error correction prescription however can be described in a unified way by means of the so called stabiliser formalism [22], [23]. The basic idea is to describe quantum states in terms of operators. Given a state $\ket{\psi}$, one can say it is being stabilised by some operator $K$ if that state is an $+1$ eigenstate of $K$ namely $K\ket{\psi}=\ket{\psi}$. A multi-qubit state will be described in an operatorial sense by analysing the group properties of the multi-qubit operators acting as stabilisers. Given the Pauli group for $N$-qubits $\mathcal{P}_{N}$, an $N$-qubit stabiliser state is defined by the $N$ generators of an Abelian subgroup $\mathcal{G}$ of the $N$-qubit Pauli group that satisfies 
\begin{equation}
\mathcal{G}=\{K^{i}\,|\,K^{i}\ket{\psi}=\ket{\psi},\; [K^{i},K^{j}]=0,\; \forall (i,j)\}\subset \mathcal{P}_{N}
\end{equation}
A given state $\ket{\psi}_{N}$ can be defined by specifying the generators of the stabiliser group. Each stabiliser operation squares to the identity. The use of stabiliser operators to describe quantum error correction codes allows us to see what logical operations can be applied directly to the encoded data. The preparation of logical state is based on the fact that valid codeword states are defined as simultaneous $+1$ eigenstates for each of the generators of the stabiliser group. Therefore it will be required to project our qubits into eigenstates of each of these operators. With the arbitrary input state $\ket{\psi}_{I}$ given, an ancilla initialised in the $\ket{0}$ state is used as a control qubit for the unitary and Hermitian operation $U$ performed on $\ket{\psi}_{I}$. A Hadamard gate is applied on the ancilla state and then it is coupled by means of the operation $U$ to our state $\ket{\psi}_{I}$. After inserting another Hadamard gate for the ancilla qubit, the state of the system will be
\begin{equation}
\ket{\psi}_{F}=\frac{1}{2}(\ket{\psi}_{I}+U\ket{\psi}_{I})\ket{0}+\frac{1}{2}(\ket{\psi}_{I}-U\ket{\psi}_{I})\ket{1}
\end{equation}
We now measure the ancilla qubit in the computational basis. If the result is $\ket{0}$ then the input state becomes 
\begin{equation}
\ket{\psi}_{F}=\ket{\psi}_{I}+U\ket{\psi}_{I}
\end{equation}
while if the measured outcome of the ancilla is $\ket{1}$ then the input state becomes 
\begin{equation}
\ket{\psi}_{F}=\ket{\psi}_{I}-U\ket{\psi}_{I}
\end{equation}
Therefore this circuit projects onto the $\pm 1$ eigenstates of $U$. In order to project onto the positive eigenstate we read the measurement and decide whether to apply a gate that will project on the positive eigenstate, call it $Z$. To generalise this circuit for the situation in which we have several stabiliser operators we simply connect each stabiliser gate to the corresponding ancilla and measure the outcomes of all ancillas before projecting if necessary. In this way we will have projections onto the common eigenstates of the stabiliser operators. Quantum codes can be characterised by means of the number of physical qubits $(n)$ encoding a certain number of logical qubit $(k)$ with the associated distance between basis states $(d)$ as $[[n,k,d]]$. Then we may consider for the sake of example the quantum code $[[7,1,3]]$ which can correct $t=1$ error. The code defines one single logical qubit and hence must contain two meaningful logical code states $\ket{0}$ and $\ket{1}$ which are basis states for the code and can be written in a state vector notation for physical qubits as 
\begin{widetext}
\begin{equation}
\begin{array}{c}
\ket{0}_{L}=\frac{1}{8}(\ket{0000000}+\ket{1010101}+\ket{0110011}+\ket{1100110}+\ket{0001111}+\ket{1011010}+\ket{0111100}+\ket{1101001})\\
\\
\ket{1}_{L}=\frac{1}{8}(\ket{1111111}+\ket{0101010}+\ket{1001100}+\ket{0011001}+\ket{1110000}+\ket{0100101}+\ket{1000011}+\ket{0010110})\\
\end{array}
\end{equation}
\end{widetext}
But in the case in which we work on a logical state encoded as a 7-qubits physical state, the total dimension of the Hilbert space must be $2^{7}$. However, the logically encoded state will only require a 2-dimensional subspace spanned by the states above. Stabiliser groups and their operators make such a reduction visible. For a $7$ qubits code we have six stabiliser operators. These will reduce the dimension of the code subspace as expected to $2^{7-6}=2^{1}=2$ which is the dimension of the logical qubit. Error correction for stabiliser codes is an extension of the state preparation prescription. Assume that on the encoded state 
\begin{equation}
\alpha\ket{0}+\beta\ket{1}\rightarrow\alpha\ket{0}_{L}+\beta\ket{1}_{L}=\ket{\psi}_{L}
\end{equation}
an error occurs at the level of an encoding qubit. This error is described by the operator $E$ acting over the $N$ physical qubits of the logical state. The erred state will then be
\begin{equation}
K^{i}E\ket{\psi}_{L}=(-1)^{m}EK^{i}\ket{\psi}_{L}=(-1)^{m}E\ket{\psi}_{L}
\end{equation}
The parameter $m$ is equal to zero if the error and the stabiliser commute and is equal to $1$ if they anti-commute. The error procedure implies a sequential measurement of each of the code stabilisers. If the error operator commutes with the stabiliser the state remains a $+1$ eigenstate of $K^{i}$, while if the error operator anti-commutes with the stabiliser then the logical state is flipped to a $-1$ eigenstate of $K^{i}$. The procedure of error correction is equivalent with that of state preparation. Since an error free state is already a $+1$ eigenstate of the stabilisers, the anti-commuting errors with any of the stabilisers will flip the relevant eigenstate and therefore when we measure the parity of these stabilisers we will obtain $\ket{1}$. For the $[[7,1,3]]$ code if the error operator is $E=X_{i}$ with $i=1,...,7$ encoding a bit flip on any one single qubit of the $7$ physical qubits, then, no matter where such bit flip would occur, $E$ will anti-commute with a unique combination of $K^{4}$, $K^{5}$, and $K^{6}$. After measuring these three operators we will obtain information about whether and where the $X_{i}$ error occurred. If $E=Z_{i}$ the error operator will anti-commute with a unique combination of $K^{1}$, $K^{2}$, and $K^{3}$ and will give us information about the $Z$ error. 
This example, based on the $[[7,1,3]]$ code, while certainly limited, is useful in understanding how the main idea of this article will be developed in the case of holographic quantum error correction codes and their topological properties. Indeed, stabiliser operators may be used to generally specify error correction codes and to reduce the dimensionality of the physical Hilbert space down to the subspace that encodes our logical states. 
Another source of errors is the actual loss of physical qubits. The loss of, say, a photon is assumed to be equivalent to measuring the photon in a basis, say $\{\ket{0},\ket{1}\}$ without knowing the answer. Such ignorance results in a possible logical bit-flip error on the encoded state, and hence the problem will be how to protect against logical bit flip errors. We already saw that the 3-qubit code allows us to obtain this type of correction. The important part is to encode the states into a redundancy code where an arbitrary logical state $\ket{\psi}_{L}$ is now given by 
\begin{equation}
\ket{\psi}_{L}=\alpha\ket{0}_{1}^{N}\ket{0}_{2}^{N}...\ket{0}_{q}^{N}+\beta \ket{1}_{1}^{N}\ket{1}_{2}^{N}...\ket{1}_{q}^{N}
\end{equation}
where $\ket{0}^{N}$ and $\ket{1}^{N}$ are the so called parity encoded states. The general parity encoding for a logical qubit is an N-photon GHZ state in the conjugate basis [25]
\begin{equation}
\begin{array}{c}
\ket{0}_{L}^{N}=\frac{1}{\sqrt{2}}(\ket{+}^{\otimes N}+\ket{-}^{\otimes N})\\
\\
\ket{1}_{L}^{N}=\frac{1}{\sqrt{2}}(\ket{+}^{\otimes N}-\ket{-}^{\otimes N})\\
\end{array}
\end{equation}
where $\ket{\pm}=(\ket{0}\pm \ket{1})/2$. This type of encoding is useful because measuring any qubit in the $\{\ket{0},\ket{1}\}$ basis removes it from the state, with the result state being reduced
\begin{equation}
\begin{array}{c}
P_{0,N}\ket{0}_{L}^{N}=(I_{N}+Z_{N})\ket{0}_{L}^{N}=\\
\\
=\frac{1}{\sqrt{2}}(\ket{+}^{N-1}+\ket{-}^{N-1})\ket{0}_{N}=\ket{0}_{L}^{N-1}\ket{0}_{N}\\
\\
P_{1,N}\ket{0}_{L}^{N}=(I_{N}-Z_{N})\ket{0}_{L}^{N}=\\
\\
=\frac{1}{\sqrt{2}}(\ket{+}^{N-1}-\ket{-}^{N-1})\ket{1}_{N}=\ket{1}_{L}^{N-1}\ket{1}_{N}\\
\end{array}
\end{equation}
where $P_{i,N}$ are the projectors corresponding to the measurement in the $\ket{0}\ket{1}$ basis of the $N$-th qubit. 
Such encoding protects against the loss of qubits because it first encodes the system into a code structure that allows for the removal of qubits without eliminating the computational state and then it protects against logical errors induced by loss events. The basic idea is that this prescription maps errors un-correctable by standard error correction codes to errors that are correctable [25]. 

The extension of degrees of freedom in double field theory is required in order to encode global information regarding the phenomena that depart from the point like interpretation of elementary particles. Winding modes being strictly stringy objects will have to be represented through the doubling of the coordinates in the bulk space and their introduction will require a modification in the way logical states may be represented. The gauge invariance (a.k.a. redundancy) of the doubled field theory will incorporate additional transformations which may be interpreted in terms of quantum error correction codes. Their impact will be made clear later on in this article. 

\section{Holography as error correction}
In order to exploit holography with double field theory in the bulk as a quantum error correction code, we need to better understand how holography may be interpreted as an error correction code to begin with. Double field theory adds additional information in this context as it tries to incorporate string theoretical phenomena in effective field theories at lower energies. Looking at the holographic principle from the perspective of quantum error correction codes helps us better understand non-localities in the bulk. We expect them to exist due to the extended nature of strings, however, they are usually not manifest as the bulk boundary duality is best understood in the context where bulk physics is described by classical gravity. Introducing double field theory in the bulk makes T-duality manifest and T-duality relates not only distinct geometries, but also distinct topologies. Moreover, the correspondence between non-trivial topology and entanglement [4] shows that T-duality may play the role of  a change of the factorisation of the algebra describing the total quantum state. It is known that depending on the factorisation considered, a quantum state may appear either as entangled or as separable. For pure states we can switch between separability and entanglement in a unitary fashion. For mixed states however, we need some minimal amount of mixedness [15]. Incorporation of T-duality in the bulk is therefore crucial to the interpretation of the bulk-boundary duality as a quantum error correction code, as such codes rely on the existence of entanglement. Without a clear understanding of the topology changing phenomena occurring in the bulk, the quantum error correction code interpretation is not complete. The emergence of bulk locality and its relation to quantum error correction has been mentioned in [7] where it has been shown that all the bulk notions such as the Bogoliubov transformation, the localisation in the radial direction, and even the holographic entropy bound have natural boundary conformal field explanations in terms of quantum error corrections. Therefore, it is worthwhile exploring the interpretation of holography in terms of quantum error correction codes before we go further to understanding how T-duality and its associated topological uncertainty may impact such an interpretation. 
As mentioned in [7] but already well known to the holographic community, it is still a mystery how bulk locality emerges, even in an approximate way. Near the boundary it is quite clear that the relation 
\begin{equation}
\lim_{r\rightarrow\infty}r^{\Delta}\phi(r,x)=\mathcal{O}(x)
\end{equation}
remains valid, where we have considered the limiting values of a bulk field $\phi$ and a conformal field theory operator $\mathcal{O}$. A dictionary based on this relation will manifestly respect locality in the $x$ direction simply because the conformal field theory does so too. Moving in the radial direction, such an approximate locality is less obvious. A local operator in the centre of the bulk is expected to commute with every local operator at the boundary given a fixed time slice containing that particular bulk operator. However, it is known that any operator that commutes with all local operators at a fixed time must be proportional to the identity. Because of this, bulk locality cannot be respected within the conformal field theory at the level of the algebra of operators [7]. Of course, together with the authors of [7], we may ask in what sense it is respected? The answer of [7] is to analyse the problem of bulk locality in terms of the stability of the bulk phenomena to errors in the boundary theory. The deeper into the bulk a process occurs the more resilient it will be to local errors. The radial direction in the bulk is seen from the perspective of the CFT as a measure of how well the CFT representations of the phenomena occurring within the bulk are protected from local erasures. The holographic principle appears as an upper bound on the amount of information that can be protected from erasures. It is important to understand that most error protection or correction codes add supplemental qubits into the description therefore increasing the total number of information and the required entanglement. While usual quantum field theories in the bulk would regain locality at least in some approximate way, the natural result, incorporating T-duality, will be manifestly non-local and hence using double field theory in the bulk will give a better insight into the nature of such non-localities. The idea that truncated subalgebras of bulk observables are relevant in the analysis has been explored both in [7] and in [16]. Such factorisation can be obtained in the context of double field theory as in this case, the strong constraint leads to the restoration of the non-stringy degrees of freedom. It must however be underlined that even when the strong constraint is employed, the stringy nature still remains manifest at least through the fact that the resulting theory may be defined on non-geometric backgrounds that couldn't be obtained without T-duality. Given the AdS space and a metric having the asymptotic form 
\begin{equation}
ds^{2}\sim -(r^{2}+1)dt^{2}+\frac{dr^{2}}{r^{2}+1}+r^{2}d\Omega_{d-1}^{2}
\end{equation}
we can identify the conformal field theory that is holographically dual to this system as living on the $S^{d-1}\times \mathbb{R}$. The time direction is given by $\mathbb{R}$. The Hilbert space is given by the configurations of the fields on the $d-1$ dimensional sphere $S^{d-1}$. In the usual context of a field theory in the bulk, we can construct CFT operators for the boundary which obey the bulk equations of motion. Following reference [7] we assume that interactions in the bulk are suppressed as powers of $\frac{1}{N}$. The bulk field $\phi(x)$ will be represented as 
\begin{equation}
\phi(x)=\int_{S^{d-1}\times \mathbb{R}}dY K(x;Y)\mathcal{O}(Y)
\end{equation}
This integral is performed over the conformal boundary and $K(x;Y)$ is a smearing function which obeys the bulk equations of motion for the $x$ index and as $x$ approaches the boundary it yields the boundary limit equation. One does not naturally expect that such operators have the desired commutation relations in the bulk [7]. The expected commutations are being recovered in the perturbative domain within low point correlation functions [23] but it is expected that they would break down in states with enough excitations. When we have a representation of the bulk field $\phi(x)$ as in the above expression, it is possible to use the conformal field theoretical hamiltonian in order to express all operators $\mathcal{O}(Y)$ in terms of Heisenberg picture fields on a single Cauchy surface in the CFT [7]. If we then take $x$ to be near the boundary but not yet on it, the single time CFT representation of the bulk field still involves operators with support all over the single Cauchy surface in the CFT, $\Sigma$. A representation has been found whose boundary support becomes smaller as the operator approaches the boundary. This is exactly the AdS-Rindler representation discussed in [7]. In the AdS-Rindler construction the same bulk field operator $\phi(x)$ lies in multiple causal wedges. Its representation then can exist on different distinct regions of $\Sigma$. Given any bulk field operator $\phi(x)$ and any CFT local operator $\mathcal{O}(Y)$ chosen such that $x$ and $Y$ are spacelike separated, it is possible to choose a causal wedge $\mathcal{W}_{C}[A]$ such that $\mathcal{O}(Y)$ lies in the complement of $A$ in the surface $\Sigma$. But we assume locality in the boundary CFT. Therefore $\mathcal{O}(Y)$ must commute with the representation of $\phi(x)$ in that wedge. But no non-trivial operator in the boundary conformal field theory can commute with all local CFT operators on $\Sigma$. Therefore, the representations of the bulk field in the various wedges  cannot be the same operator on the CFT Hilbert space. To better understand this apparent contradiction, the theory of quantum error corrections has been invoked by [7].

\section{Holographic error correction codes with tensor networks}
It is well known that the holographic principle implies that a theory of gravity in a bulk space is dual to a quantum field theory on a boundary. In the AdS/CFT context this is translated into a duality between a weakly coupled gravity in the bulk and a strongly coupled conformal field theory on the boundary. In order for this duality to be meaningful we need to relate the bulk operators to boundary operators. This mapping has however some surprising aspects. Of course, while introducing the bulk space we identify a new radial dimension from the boundary towards the bulk. It has been shown in [27], [28] that such a radial dimension can be seen as a renormalisation group scale. We can see the radial coordinate of a spacetime with asymptotically AdS geometry as a flow parameter of the boundary field theory. Recent research has completed the idea that geometry, seen as an emergent property, is related to quantum entanglement in the sense that the geometry in the bulk can be expressed in terms of the entanglement structure of the boundary quantum field theory. Describing quantum field theory by non-geometric means is highly complicated, therefore its connection to geometry offered a new understanding of various quantum field theoretical phenomena [29], [30]. The connection to geometry has been made clear already by the introduction of the Ryu-Takayanagi formula [31] and its covariant counterpart [32]. This formula is known to acquire corrections by various local and non-local terms. Such terms, derived also in [33] can be seen by means of so called holographic quantum correction codes. Such codes not only demonstrate the idea that entanglement is a source from where geometry emerges, but also allows us to better understand various prescriptions of the AdS/CFT dictionary. Such a construction is based on a tensor network which is expressed in terms of polygons that are uniformly tiling the bulk space. The terminology here will become that of quantum information theory, and hence we will have physical quantum information units encoding the information of logical quantum states. The physical variables associated to the quantum code will be on the boundary while the logical operators reside in the bulk. Holographic codes allow us to explicitly compute the mapping between boundary and bulk and hence to derive the dictionary of AdS/CFT. In essence, local operators in the bulk theory are being mapped into non-local operators of the bulk. This allows us to connect bulk geometry to the entanglement structure of the boundary quantum field theory. The bulk Hilbert space or the code space is a proper subspace of the boundary Hilbert space preserved by the bulk operators. The idea of reconstructing bulk operators on the boundary is based on the AdS-Rindler reconstruction described in the introduction. The ambiguity of the reconstruction prescription has been resolved by making use of some form of redundancy. Either we considered the highly different boundary operators as being different physical representations of the same type of action on the code subspace, or we considered the distinction in the boundary theory as given by the redundant description given by gauge invariance. However, the usual gauge invariance on the boundary seemed problematic. Extending it to the gauge invariance of double field theory however may have certain benefits.  The fact that operators residing in the causal wedge of a certain boundary region $A$ can be reconstructed on the boundary is well known. However, if the the boundary region $A$ is a union of two or more disconnected components, then the domain of the bulk from where operators can be reconstructed on the boundary region $A$ increases. This introduces the so called entanglement wedge $E(A)$ which may extend further into the bulk and from which bulk operators may be reconstructed on the disconnected region $A$. Moreover, entangled pairs in the bulk with one of the members inside the entanglement wedge of the region $A$ and the other outside, will contribute to the entropy of the region $A$ and hence to the entanglement shared by $A$ and its complement $\bar{A}$. This means that there should be operators in $A$ capable of detecting the member of the pair inside the entanglement wedge $E(A)$. Given the general entropy formula
\begin{equation}
S_{A}=\frac{A}{4G_{N}}+S_{bulk}(\rho_{E(A)})+\frac{\delta A}{4 G_{N}}+...
\end{equation}
the first term is the leading Ryu-Takayanagi term, which is local on the entangling surface and state independent, the second term is the bulk entropy in the entanglement wedge $E(A)$ defined by the Ryu-Takayanagi minimal surface and therefore generally non-local and non-linearly dependent on the bulk state, and the third term is an additional quantum correction to the Ryu-Takayanagi area which is both local on the minimal surface and linear in the bulk state. The last term appears to also originate from a quantum error correction code which is based on the operators $\mathcal{O}$ associated precisely to the boundary between the entanglement wedge of the boundary area $A$ and that of its complement $\bar{A}$ [34]. Once the form of this correction was known, it has been noticed that it can be derived from holographic quantum error correction codes as well [26]. The operators $\mathcal{O}$ must be reconstructible from both $A$ and $\bar{A}$ and hence the operator itself must lie in the centre of either reconstructed algebra. Terms like the first and the third in the entropy formula are related to aspects of the code derived from the values of the operators in this centre [26]. The minimal area computation in the Ryu Takayanagi formalism is translated into the calculation of the so called "greedy" surface [26] in the context of holographic codes. MERA tensor networks also realise a hyperbolic geometry and entropy bounds as those found in holographic discussions. The description of such tensor networks rely on the so called perfect tensors which arise in the expansion of a pure state describing $2n$ $v$-dimensional spins in a suitable basis
\begin{equation}
\ket{\psi}=\sum_{a_{1},...,a_{2n}}T_{a_{1}...a_{2n}}\ket{a_{1}a_{2}...a_{2n}}
\end{equation}
We call the tensor $T$ perfect if the state is maximally entangled across any bipartition cut of the set of $2n$ spins into two sets of $n$ spins. This expansion can be rewritten (by the maps of bra into ket states) in various forms and for the 6 spins code we have the operators
\begin{equation}
\begin{array}{c}
\sum_{a_{1},...,a_{6}}T_{a_{1}...a_{6}}\ket{a_{4}a_{5}a_{6}}\bra{a_{1}a_{2}a_{3}}\\
\\
\sum_{a_{1},...,a_{6}}T_{a_{1}...a_{6}}\ket{a_{3}a_{4}a_{5}a_{6}}\bra{a_{1}a_{2}}\\
\\
\sum_{a_{1},...,a_{6}}T_{a_{1}...a_{6}}\ket{a_{2}a_{3}a_{4}a_{5}a_{6}}\bra{a_{1}}\\
\end{array}
\end{equation}
These expansions form isometric encoding maps which will encode a certain number of logical qubits into the emerging physical qubits. Holographic quantum error correction codes are then implemented by contracting perfect tensors taking into account the geometry of the bulk space (in this case hyperbolic) and its tiling by corresponding polygons. The uncontracted indices are particularly important in the description of gauge fields inside the bulk and hence will become important for the extension towards double field theory. In a holographic code there are two types of uncontracted indices, namely the bulk indices and the boundary indices. All other indices are contracted between tensors arising on different layers of tilings. The bulk and the boundary indices are however not separated. Because the code is essentially an isometric embedding of the bulk Hilbert space into the boundary Hilbert space, the two indices are related. Each polygon provides an isometry from incoming and bulk indices to outgoing indices. 

\section{Bulk gauge fields in the holographic tensor networks}
The implementation of bulk fields in the holographic tensor network has been discussed in [26]. There, a generalisation of the holographic quantum error correction code for bulk gauge fields is presented. As most of the novel aspects of double field theory are revealed in the generalised gauge transformations they introduce, understanding briefly how gauge fields and gauge invariance are implemented in a holographic quantum error correction code seems essential. New degrees of freedom on the links of the holographic tensor network are being introduced to that end, and additional connections to further copies of the holographic code are implemented by suitable isometries. In the case of double field theory such degrees of freedom on the links must be extended even further, considering the topological properties of T-duality. In the non-doubled case boundary regions allow the reconstruction of bulk algebras with central elements in the interior edges of the entanglement edge. In the case of double field theory bulk algebras are further extended leading to new error correction codes, previously unavailable. A tensor network has an upper bound for the amount of entanglement the state described by it can have, and this is based on the minimal cut dividing the network. This upper bound is saturated for connected regions in certain classes of holographic states. The case of planar graphs with non-positive curvature has been described in [17]. In order for a circuit interpretation of a network of perfect tensors to be valid, according to [17] it must satisfy three criteria. The first is the covering criterium, namely that to each edge (contracted or uncontracted index) is assigned a directionality. This condition is required in order to interpret the direction in which each tensor in the network processes information and hence to meaningfully define the input and the output indices. The second condition is the so called flow condition, which implies that each tensor has an equal number of incoming and outgoing indices. This is required for the interpretation that every tensor is a unitary gate. The last condition, namely the acyclicity condition is however rather special. It is already noted in [17] that this condition is non-local and is demanded so that the order of the application of the operations in the network to be consistent. Inconsistencies in this interpretation would be the presence of a closed time-like curve in the circuit picture. The assumptions made in [17] require for the graph to be, first, a planar embedding, namely the tensor network to be laid out in a planar form, the boundary of the network being a simple boundary of the embedding. Second, the tensors are required to be perfect, having an even number of legs and being unitary along any balanced distribution of legs. Finally, the network was expected to represent an AdS bulk and hence corresponding to a network equivalent of the AdS negative curvature. This implies that the distance function between two nodes of the network has no local maxima away from the boundary. Thinking in terms of the acyclic condition, which is a non-local property, it has been shown in [17] that the presence of a cycle implies the existence of an interior local maximum for the labelling. The proof goes, according to [17] as follows. Let there be a cycle $C$ in the construction of the tensor network. The node label values immediately in the interior of the loop will be larger or smaller (depending on the orientation of $C$) than those immediate to the exteriors. In the case of $C$ counterclockwise we may chose a node in the interior of $C$ with the lowest possible label. In the case of this note, the label is smaller than those of all its neighbours including those in the exterior of $C$, which means it contradicts the assumption that it is defined based on the graph distance function and its properties. In the clockwise case, we can chose a node in the interior of $C$ with the largest possible label. In this case it represents an interior maximum for the distance function and hence the surface homeomorphic to the disc cannot be negatively curved. Fascinatingly enough, precisely this acyclicity condition cannot uphold in the case of T-duality in the bulk. But it is well known from [35] and [36] that the T-dual of the AdS spacetime is the de-Sitter spacetime and hence abandoning the acyclic condition introduces into our network the cosmologically relevant de-Sitter space. In order for this to become clear let me first follow the results of ref. [26] regarding the inclusion of gauge fields in the tensor network representation, so that in the next chapter I can bring plausibility arguments for the statement above. In the case of lattice gauge theories, there appears the requirement of additional degrees of freedom on the links of the discrete graph model in order to describe the associated gauge fields. The various holonomies arise as paths through the lattice and the Gauss constraint provide a valid gauge interpretation. In the context of the holographic code in order to treat bulk gauge theories we need to introduce additional degrees of freedom on the links of the tensor network corresponding to the pentagon code. The tensor associated to the pentagon tiling has a total of six indices, five of them being associated to the network and away from the boundary they are connected to nearby tensors. Every such tensor also has an uncontracted index associated with local bulk degree of freedom. When $T$ is a perfect tensor, it describes an isometry from any three legs to the others, then an operator $\mathcal{O}$ acting on any bulk input may be transported along three of the output legs to the three neighbouring tensors. This procedure together with the negative curvature assumption allows us to transport local bulk operators up to the boundary because each tensor has at least three legs pointing towards the boundary. The additional degrees of freedom modelling bulk gauge fields can be introduced by adding a three index tensor $G_{ijk}$
To keep the connection to the bulk, one adds the new tensor to the bulk index and connects it to another bulk index leaving the third index for additional input. In this way one merges two bulk indices into one single index. This implies contracting the new tensor $G$ with a pair of neighbouring bulk inputs as described in [26]. 

\section{Double field theory in the bulk and holographic tensor networks}
The AdS/CFT holographic correspondence appears to be the best tool of understanding non-perturbative quantum gravity. Its connection to the quantum error correction codes has been recently made manifest in [7]. The quantum error correction interpretation of holography allows us to further expand both our understanding of quantum information processing and of holography. However, AdS/CFT is somehow restricted in its applicability, as we only now start to understand how it may generalise to different quantum field theories on the boundary and new types of geometry in the bulk. In its modern interpretation, the holographic duality states that the entanglement structure (and the quantum error correcting properties) of a quantum field theory is to be interpreted as defining the geometric properties of the bulk spacetime. As far as it is known now, the correspondence connects local operators supported deep inside the bulk to highly non-local operators on the boundary. This dictionary, interpreted as the encoding map of a quantum error correcting code allows us to see bulk local operators as the logical counterparts of the physical boundary theory. The logical operators map the code subspace $\mathcal{H}_{C}$ to itself, insuring the protection of the logical information against erasures of portions of the boundary. This idea allows us to finally see the relationship between the emergence of bulk geometry and the structure of the entanglement on the boundary theory. It was shown in [38] that codes can have holographic properties even when the underlying bulk geometry does not have negative curvature. Holographic codes have been analysed in the hope of providing new insights for quantum computing architectures [37] and their description in terms of operator algebra quantum error correction has been established [39], [40]. As noted also in [37] the understanding of the bulk both in terms of physics and geometry is insufficient. While AdS/CFT is very useful, understanding the bulk in more general contexts will reveal several aspects of quantum cosmology, quantum field theory, and quantum information. Also, [37] noticed that holographic codes are locally correctable provided that the bulk geometry is negatively curved in the asymptotic limit, but is not locally correctable for asymptotic flat or positive curvature. This may result in non-local physics on the boundary in the flat and positively curved cases. When approaching the boundary, it is interesting to consider therefore certain string theoretical effects which originate from their finite size. T-duality can be associated with topology changing effects and the field theoretical interpretation that makes T-duality manifest, namely double field theory, features properties that cannot be encoded in strictly local quantum field theories. Indeed string theory has access to both globally and locally non-geometric backgrounds where non-associative phenomena may occur. Obviously, the entanglement structure of such objects will be particularly complicated and I do not intend to explore it here directly. Instead of doing this, I will formulate a strategy for relating gauge symmetry in double field theories to particular features of holographic quantum error correction codes. Geometrical properties inside the bulk are related to the structure of quantum entanglement in the boundary, but then, to what can non-geometrical properties associated to double field theory and string theoretical T-duality in the bulk be related on the boundary? Can we establish a new form of quantum error correction on the boundary by thinking in terms of non-geometric properties in the bulk? It appears that the answer to both questions is in the positive. As is known, local operators in the bulk can be seen as logical operators acting on the code subspace. For holographic codes, the fact that a subsystem of the physical Hilbert space $\mathcal{H}$ is correctable with respect to a logical sub-algebra can be interpreted according to [37] in terms of a question about the bulk geometry. This can be seen from the perspective of the entanglement wedge hypothesis which provides us with the largest bulk region with a logical sub-algebra that can be represented on a given boundary region. In the case of double field theory and string geometry, several non Riemannian phenomena can occur in the bulk, leading to different ways in which manifolds can be patched together. If T-duality is being employed as a transition function, we find a new class of backgrounds which present a non-trivial dependence on the dual coordinates that are conjugate to the string winding number. As has been shown in the previous section, to introduce the degrees of freedom associated to gauge fields on links, the bulk indices of the code tensors must be linked with a tensor called $G_{ijk}$ whose role in the network is to add the degrees of freedom required. To keep the input from the bulk space into the network tensors $T$, this tensor has been extended to a six-fold tensor product 
\begin{equation}
T\rightarrow \bigotimes_{m=1}^{6}T
\end{equation}
resulting that these units must be connected as in a pentagonal tiling of the hyperbolic disk. Each factor of the resulting tensor product was called a copy of the network. The first copy was considered to be the original network and the other copies were contracted with the three-legged tensor $G$. In this way the network will have six bulk input legs at each vertex. Five of these can be turned into inputs for the edges. Take one edge in the interior of the disk and one input leg from each of the two vertices it connects. These legs will be contracted with two of the three legs of the tensor $G$. These two legs of $G$ will be considered output legs and the remaining one will be an input leg associated with the current edge under consideration. This uses up five of the bulk legs at each vertex leaving the last one as a normal bulk input leg at each vertex. This prescription adds one $G$ for every two bulk legs and implicitly for two tensors $T$. This construction is as the one described in [26]. In order to introduce T-duality and the effects related to double field theory, certain adaptations must be included. The most intuitive modification that allows the doubling of the fields in the bulk is to increase the number of degrees of freedom associated to the tensor $G$, for example by adding a new index. This alters the connectivity of the bulk in the sense that now, two output legs of $G$ will be connected to the input legs of the vertices while another two legs of the tensor $G$ will play the role of input legs. This will connect the various layers of the bulk tensor network and will provide us with new topological structures not available before. In particular, the acyclicity condition must be altered in order to take into account effects that are related to T-duality. In particular, being a non-local property, it can be interpreted as a topological structure of the bulk spacetime. Topologically non-trivial bulk surfaces are not unexpected, particularly situations manifesting entanglement over spacelike separated regions, this being the main idea behind the ER-EPR duality [41]. The effect of this extension however has another interesting consequence. As shown previously, the condition of acyclicity not only reassures us that no closed timelike curves are possible, but also keeps the distance function from having a maximum away from the boundary. As doubling the number of bulk degrees of freedom alters this property, it becomes obvious that such a property can no longer be maintained, leading us to conclude that we cannot assume that the negative curvature assumed in the beginning can be preserved. Therefore, it appears that double field theory in the bulk, by its ability of manifestly implementing T-duality leads to positive curvature and hence to the emergence of a de-Sitter spacetime. The connections of such an observation with cosmological data remains to be discussed in a future article. Until then, following ref. [37], it will be relevant to move to the operator algebra quantum error correction code interpretation in order to see the way in which gauge freedom manifests itself in the bulk and how the extended gauge freedom of double field theory affects the standard holographic interpretation. In the simplest case, let there be $H$ a finite dimensional Hilbert space and an associated complex vector space of linear operators acting on this Hilbert space. This complex vector space forms an algebra $\mathcal{A}$. We can identify a subspace of this Hilbert space which can be written as a product of two tensor factors 
\begin{equation}
H\supseteq \bigoplus_{\alpha}H_{\alpha}\otimes H_{\bar{\alpha}}
\end{equation}
The algebra then can be written as 
\begin{equation}
\mathcal{A}=\bigoplus_{\alpha}\mathcal{M}_{\alpha}\otimes I_{\bar{\alpha}}
\end{equation}
We define the commutant of $\mathcal{A}$ as the algebra $\mathcal{A}'$ which contains all operators acting on the Hilbert space $H$ which commute with all the operators of the algebra $\mathcal{A}$. Then the commutant algebra can be written as 
\begin{equation}
\mathcal{A}'=\bigoplus_{\alpha}I_{\alpha}\otimes \mathcal{M}_{\bar{\alpha}}
\end{equation}
The original algebra and its commutant share the same centre $Z(\mathcal{A})$ which contains elements of the form 
\begin{equation}
\bigoplus_{\alpha}m_{\alpha}I_{\alpha}\otimes I_{\bar{\alpha}}
\end{equation}
where $I$ represents the identity matrix on the respective subspace. 
The above algebras describe both the classical and quantum aspects of the information related to a system. The operators in the common centre describe the classical data, like the area operator associated to the minimal area in the Ryu-Takayanagi formula, while $\mathcal{M}_{\alpha}$ describes the quantum data. In order to analyse error correction properties of holographic codes it is useful to consider the code subspace of the Hilbert space associated to a low energy domain of the boundary field theory. Here, the algebras $\mathcal{A}$ and $\mathcal{A}'$ contain logical operators that preserve the code subspace. In the case of a single summand in the decomposition of the Hilbert space, with $\mathcal{M}_{\bar{\alpha}}$ non-trivial, the algebra $\mathcal{A}$ represents the algebra of logical operators in a subsystem code, while the code subspace can be decomposed in two sides, 
\begin{equation}
H_{C}=H_{\alpha}\otimes H_{\bar{\alpha}}
\end{equation}
the first representing the protected tensor part, $H_{\alpha}$, while the other represents a gauge part, $H_{\bar{\alpha}}$. The algebra $\mathcal{A}$ acts only on the protected part, while by means of the holographic duality, the gravitational bulk system has an emergent gauge symmetry. In double field theory this gauge symmetry is extended in a non-trivial way, by introducing additional components related to the winding modes. Basically, as stated in the second chapter, the doubled gauge parameter is extended as $\xi^{M}=(\tilde{\lambda}_{i}, \lambda^{i})$. The doubling has as effect the extension of the bulk algebra to one with non-trivial commutation relations linking the extended winding modes and the associated coordinates. The basis of the doubled space are the so called left and right moving modes that now can be formed given the non-local effects provided by T-duality. The idea of forming the left-invariant and right-invariant forms for both left and right moving modes implies the doubling of the associated algebra leading to elements belonging to the direct product of these. As we have these, let us consider the algebra generated by the operations $G_{I}$ obeying commutation relations of the form 
\begin{equation}
\begin{array}{ccc}
[G_{I},G_{J}]=i\cdot f_{IJ}^{\;\;\;\;K}G_{K}, & Tr(G_{I}G_{J})=\eta_{IJ}, & \det(\eta_{IJ})\neq 0\\
\end{array}
\end{equation}
where $\eta_{IJ}$ is a non-degenerate metric doubled metric. In our context of the $AdS$ bulk spacetime, we need a doubling of this space into left and right components. Representing this in the form of groups, the doubled $AdS$ group will be $SO(D,D+1)$ according to [42]. The doubled $AdS$ algebra is generated by the doubled momenta, and the doubled Lorentz generators, and it results in the emergence of a left / right mixed index. We observe [42] that the left moving mode will exist in the $AdS$ space while the right moving one in the de-Sitter space. The details of the group representation of the doubled bulk space as well as the emergence of the de-Sitter components have been presented in [42] and I will not insist upon them here. The relevant aspect however is first, that the explicit introduction of T-duality in the bulk by means of field doubling leads to modes belonging to the de-Sitter component, and second, that such an extension can be encoded by extending the gauge component in the bulk corresponding to a boundary term that originated in a particular map of the boundary code subspace. To see how the boundary theory implements such a doubling in terms of quantum error code correction, let me follow again [37] to see how the error correction map can be extended. Of course, a more comprehensive discussion will have to take into account an actual string theory in the bulk, but this lies beyond the scope of the present article. Given an noise channel which can be written as acting on an operator $X$ as
\begin{equation}
\mathcal{N}(X)=\sum_{a}N_{a}^{\dagger}XN_{a}
\end{equation}
with $N_{a}$ being the Krauss operators [37], and considering the Hilbert space $H$ then quantum error correction would be a process that could reverse the effects of $\mathcal{N}$. Unless $\mathcal{N}$ is unitary, such a reversion would not be possible over the entire $H$ but we still hope to be able to reverse a subset $H_{C}\subset H$ of the original Hilbert space. Indeed considering the algebra of logical operators that act on the code space we denote the set of those linear operators that map $H$ into $H$ as $\mathcal{L}(H)$ and given the projector $P$ from $H$ to $H_{C}$, the operator $X\in \mathcal{L}(H)$ is called logical if $[X,P]=0$ and therefore $X$ maps the code space to itself
\begin{equation}
X\cdot H_{C}=XP\cdot H=PX\cdot H\subseteq H_{C}
\end{equation}
We say that the noise $\mathcal{N}$ is correctable on the code space $H_{C}$ with respect to the operator $X\in\mathcal{L}(H)$ if there exists such a recovery channel $\mathcal{R}$ that the property 
\begin{equation}
P(\mathcal{R}\cdot \mathcal{N})^{\dagger}(X)P=PXP
\end{equation}
is satisfied. In the bulk however, correctability is represented in terms of the so called entanglement wedge hypothesis which states basically that if a bulk region is included in the entanglement wedge of a boundary region, then the complementary boundary region is correctable with respect to the logical bulk subalgebra associated to the above mentioned bulk region. In the case in which the bulk space is doubled and mixing indices appear connecting anti-de-Sitter and de-Sitter algebras, the standard representation of the bulk algebra as a tensor product over bulk states 
\begin{equation}
\mathcal{A}=\bigotimes_{x}\mathcal{A}_{x}
\end{equation}
fails, in the sense that there can be no such simple decomposition as the group associated to the bulk space will include also a de-Sitter component. However, it is still possible to reconstruct a boundary theory by carefully restricting the doubled coordinates in the bulk region close to the boundary. There are several options by which such a restriction can be performed [43], [44], [45], each coming with advantages and disadvantages. The most important aspect is to keep the desirable effects of T-duality in the limit where the doubled coordinates become irrelevant. It is currently not clear what precisely the boundary encoding map associated to the mixing terms found in the bulk due to the extended gauge symmetry. Heuristically speaking, it is possible to imagine that stringy modes encoded in the bulk double field theory may have the effect of violating associativity of the operators in the boundary. Also, effects associated to the extended nature of the strings, from where double field theory extracts its stranger features, seem to be related to mixing of operators in the boundary and to a left-right symmetry which should not otherwise be present [46].
\section{Conclusions}
This article explores the connection between quantum code correctability and the geometry of the bulk in the unexplored context in which the bulk coordinates are doubled in order for string theoretical duality to become manifest. It seems plausible nowadays that for holographic codes, the correctability of a subsystem of a Hilbert space can be expressed in terms of the bulk geometry by means of the so called entanglement wedge hypothesis. Such an observation basically equates the emergence of spacetime geometry with the structure of entanglement in a non-gravitational quantum field theory. However, such an analysis does not directly take into account gauge invariance, which may play an important role in establishing the connection between quantum error code correction and geometry. Even more so, stringy effects are not directly considered and they may be the key towards even more advanced quantum codes. This article is a heuristic attempt towards the exploration of the holography / quantum error correction duality in the context in which the stringy T-duality becomes manifest. The main observation is that T-duality plays an important role in the emergence of de-Sitter geometry in the bulk which may lead to further cosmological discussions. Moreover, T-duality appears to negate the requirement for acyclicity in the bulk, a phenomenon that may lead to new interpretations of the dualities between topologically distinct manifolds in the context of the ER-EPR duality. This article wishes to point out these new directions leaving for future work, the task of going from a heuristic approach to one based on more specific examples.


\begin{thebibliography}{99}
\bibitem{1} E. Mintun, J. Polchinski, V. Rosenhaus, Phys. Rev. Lett. 115, 151601 (2015)
\bibitem{2} A. T. Patrascu, Condens. Matter, 2 (4), 33 (2017)
\bibitem{3} O. Holm, C. Hull, B. Zwiebach, JHEP 2010:08 (2010)
\bibitem{4} A. T. Patrascu, JHEP 06, 46 (2017)
\bibitem{5} J. Maldacena, L. Susskind, Fortsch. Phys. 61, pag. 781-811 (2013)
\bibitem{6} G. Aldazabal, D. Marques, C. Nunez, Class. Quantum Grav. 30, 163001 (2013)
\bibitem{7} A. Almheiri, X. Dong, D. Harlow, JHEP, 04, 163 (2015)
\bibitem{8} P. W. Shor, Phys. Rev. A 52, R2493 (1995)
\bibitem{9} M. Grassl, T. Beth, T. Pellizzari, Phys. Rev. A 56, 33 (1977)
\bibitem{10} O. Hohm, B. Zwiebach, JHEP 05, 126 (2012)
\bibitem{11} I. Jeon, K. Lee, J. H. Park, Phys. Rev. D 84, 044022 (2011)
\bibitem{12} G. Aldazabal, M. Grana, S. Iguri, M. Mayo, C. Nunez, J. A. Rosabal, JHEP 03, 093 (2016)
\bibitem{13} J. W. Lee, J. Lee, H. C. Kim, JCAP 08, 005 (2007)
\bibitem{14} P. W. Shor, Proc. of 37th FOCS, pag. 56
\bibitem{15} W. Thirring, R. A. Bertlmann, P. Kohler, H. Narnhofer, Eur. Phys. J. D 64, pag. 181 (2011)
\bibitem{16} K. Papadodimas, S. Raju, Phys. Rev. D 89, 086010 (2014)
\bibitem{17} F. Pastawski, B. Yoshida, D. Harlow, J. Preskill, JHEP 06, 149 (2015)
\bibitem{18} L. Susskind, J. of Math Phys. 36, 6377 (1995)
\bibitem{19} A. T. Patrascu, Condens. Matter, 2 (2), 13 (2017)
\bibitem{20} I. A. Batalin, G. A.Vilkovisky, Nucl. Phys. B, Vol. 234, 1, Pag. 106 (1984)
\bibitem{21} J. Gomisab, J. Paris, S. Samuel, Phys. Rep. Vol. 259, 1-2, Pag. 1 (1995)
\bibitem{22} D. Gottesman, Phys. Rev. A 54, 1862 (1996)
\bibitem{23} D. Gottesman, Ph.D. Thesis (Caltech), quant-ph/9705052 (1997)
\bibitem{24} D. Kabat, G. Lifschytz, D. A. Lowe, Phys. Rev. D 83, 106009 (2011)
\bibitem{25} S. J. Devitt, W. J. Munro, K. Nemoto, Rep. on Prog. in Phys. Vol. 76, 7 (2013)
\bibitem{26} W. Donnelly, D. Marolf, B. Michel, J. Wien, JHEP 04, 093 (2017)
\bibitem{27} M. Fukuma,  S. Matsuura, T. Sakai, Prog. of Theor. Phys., Vol. 109, 4, Pag 489 (2003)
\bibitem{28} J. de Boer, E. Verlinde, H. Verlinde, JHEP 08, 003 (2000)
\bibitem{29} P. Calabrese, J. Cardy, J. Stat. Mech: Theor. and Exp., P06002 (2004)
\bibitem{30} Y. Shi, Phys. Rev. D 70, 105001 (2004)
\bibitem{31} S. Ryu, T. Takayanagi, Phys. Rev. Lett. 96, 181602 (2006)
\bibitem{32} V. Hubeny, M. Rangamani, T. Takayanagi, JHEP 07, 062 (2007)
\bibitem{33} T. Faulkner, A. Lewkowycz, J. Maldacena, JHEP 11, 074 (2013)
\bibitem{34} D. Harlow, Commun. Math. Phys. 354, 865 (2017)
\bibitem{35} M. Hatsuda, K. Kamimura, W. Siegel, JHEP 05, 069 (2017)
\bibitem{36} C. Klimcik, JHEP 12, 051 (2002)
\bibitem{37} F. Pastawski, J. Preskill, Phys. Rev. X, 7, 021022 (2017)
\bibitem{38} P. Hayden, S. Nezami, X.-L. Qi, N. Thomas, M. Walter, Z. Yang, JHEP 11, 009 (2016)
\bibitem{39} D. W. Kribs, R. Laflamme, D. Poulin, Phys. Rev. Lett. 94, 180501 (2005)
\bibitem{40} C. Beny, A. Kempf, D. W. Kribs, Phys. Rev. A 76, 042303 (2007)
\bibitem{41} J. Maldacena, L. Susskind, Fortsch. Phys. 61, pag. 781 (2013)
\bibitem{42} M. Hatsuda, K. Kamimura, W. Siegel, JHEP 05, 069 (2017)
\bibitem{43} A. Betz, R. Blumenhagen, D. Lust, F. Rennecke, JHEP 05, 044 (2014)
\bibitem{44} K. Lee, Nucl. Phys. B, Vol. 909, pag. 429 (2016)
\bibitem{45} C. T. Ma, Phys. Rev. D 92, 066004 (2015)
\bibitem{46} A. Datta, S. Pakvasa, U. Sarkar, Phys. Lett. B, Vol. 313, 1, pag. 83 (1993)
\end{thebibliography}
\end{document}